\documentclass[twocolumn,superscriptaddress]{revtex4}
\usepackage[dvips]{graphicx}
\usepackage{amssymb,amsfonts,amsmath}
\usepackage{color}
\usepackage{eurosym}
\usepackage[sort&compress]{natbib}

\begin{document}

\title{\sffamily Evolutionary games in the multiverse}

\author{Chaitanya S. Gokhale}
\address{Emmy-Noether Group for Evolutionary Dynamics,
Department of Evolutionary Ecology,
Max-Planck-Institute for Evolutionary Biology, August-Thienemann-Stra{\ss}e 2, 24306 Pl\"{o}n, Germany}
\author{Arne Traulsen}
\email{traulsen@evolbio.mpg.de}
\address{Emmy-Noether Group for Evolutionary Dynamics,
Department of Evolutionary Ecology,
Max-Planck-Institute for Evolutionary Biology, August-Thienemann-Stra{\ss}e 2, 24306 Pl\"{o}n, Germany}

\sffamily


\begin{abstract}
Evolutionary game dynamics of two players with two strategies has been studied in great detail.
These games have been used to model many biologically relevant scenarios, ranging from social dilemmas in mammals to microbial diversity.
Some of these games may in fact take place between a number of individuals and not just between two.
Here, we address one-shot games with multiple players.
As long as we have only two strategies, many results from two player games can be generalized to multiple players.
For games with multiple players and more than two strategies, we show that statements derived for
pairwise interactions do no longer hold.
For two  player games with any number of strategies there can be at most one isolated internal equilibrium.
For any number of players $\boldsymbol{d}$ with any number of strategies $\boldsymbol{n}$, there can be at most $\boldsymbol{(d-1)^{n-1}}$ isolated internal equilibria.
Multiplayer games show a great dynamical complexity that cannot be captured based on pairwise interactions.
Our results hold for any game and can easily be applied for specific cases, e.g. public goods games or multiplayer stag hunts. 
 \end{abstract}

\maketitle

Game theory was developed in economics to describe social interactions, 
but it took the genius of John Maynard Smith and George Price to transfer this idea to biology and develop Evolutionary Game Theory \cite{maynard-smith:1973to,maynard-smith:1982to,nowak:2006bo}.
Numerous books and articles have been written since.
Typically, they begin with an introduction about evolutionary game theory and go on to describe the Prisoners Dilemma, which
is one of the most intriguing games because rational individual decisions
lead to a deviation from the social optimum. 
In an evolutionary setting, the average welfare of the population
decreases, since defection is selected over cooperation. 
How can a strategy spread that decreases the fitness of an actor, but increases the fitness of its interaction partner?  Various ways to solve such social dilemmas have been proposed \cite{nowak:2006pw,taylor:2007bb}. 
In the multiplayer version of the Prisoners Dilemma, the Public Goods Game, a number of players take part by contributing into a common pot.
Interest is added to it and then the amount is split equally amongst all, regardless of whether they have contributed or not.
Since only a fraction of one's own investment goes back to each player,
there is no incentive to deposit anything. Instead, it is tempting only to take the profits of the investments of others.
This scenario has been analyzed in a variety of contexts \cite{ostrom:1990bo,hauert:2002te}.
The evolutionary dynamics of more general multiplayer games has received considerably less attention and 
we can guess why from
 the way Hamilton put it, ``The theory of many-person games may seem to stand to that of two-person games in the relation of sea-sickness to a headache" \cite{hamilton:1975aa}.
Only recently, this topic has attracted renewed interest \cite{broom:1997aa,hauert:2006fd,pacheco:2009aa,souza:2009aa,kurokawa:2009aa,veelen:2009ma}.

As shown by Broom et al.\ \cite{broom:1997aa},
the most general form of multiplayer games, 
a straightforward generalization of the payoff matrix concept, leads to a significant increase in the complexity of the evolutionary dynamics. 
While the evolution of cooperation is an important and illustrative example, typically it does not lead to very complex dynamics. 
On the other hand, intuitive explanations for more general games are less straightforward, but only they illustrate the full dynamical complexity of multiplayer games \cite{broom:1997aa}.

To approach this complexity, we discuss evolutionary dynamics 
in finite as well as infinite populations.
For finite populations, we base our analysis on a variant of the Moran process \cite{nowak:2004pw}, 
but under weak selection, our approach is valid for a much wider range of evolutionary processes, see next section.
We begin by recalling the well studied two player two strategies scenario.
Then, we increase the number of players which results in a change in the dynamics and some basic properties of the games.
For infinitely large populations, we explore the dynamics of multiplayer games with multiple strategies and illustrate that this new domain is very different as compared to the two player situation
(see also \cite{broom:1997aa}).
We provide some general results for these multiplayer games with multiple strategies.
The two strategy case and the two player scenario are then a special case, a small part of a bigger and more complex multiverse.

\section{Model and Results}

Two player games with two strategies have been studied in detail, under different dynamics and for infinite as well as for finite population sizes.
Typically, two players meet, interact and obtain a payoff. 
The payoff is then the basis for their reproductive success and hence for the change in the composition of the population \cite{maynard-smith:1982to}.
This framework can be used for biological systems, where strategies spread by genetic reproduction, and for social systems, where strategies spread by cultural imitation.

Consider two strategies, $A$ and $B$.
We define the payoffs by $\alpha_i$ where $\alpha$ is the strategy of the focal individual and the subscript $i$ is the number of remaining players playing $A$.
For example, when an $A$ strategist meets another person playing $A$ she gets $a_1$.
She gets $a_0$ when she meets a $B$ strategist.
This leads to the payoff matrix
\begin{align}\label{eq:twobytwo}
\begin{array}{c cc}
\hline\hline
 &$A$	&	$B$\\
\hline
$A$ 	& a_1 &	a_0 
 \\
 $B$ 	&  b_1 &b_0 \\
 \hline\hline
\end{array}
\end{align}
Some of the important properties of two player games are:
\begin{itemize}

\item [(1)] {Internal equilibria}.
 When $A$ is the best reply to $B$ ($a_{0}>b_{0}$) and $B$ is the best reply to $A$ ($b_{1}>a_{1}$),  the replicator dynamics  predicts a stable coexistence of both strategies.
 Similarly, when both strategies are best replies to themselves, there is an unstable coexistence equilibrium.
A two player game with two strategies can have at most one such internal equilibrium.

\item[(2)] {Comparison of strategies}.
In a finite population, strategy $A$ will replace $B$ with a higher probability than vice versa if 
$N a_0 + (N-2) a_1
>
(N-2) b_0 +N b_1$. 
This result holds for the deterministic evolutionary dynamics discussed by Kandori et al. \cite{kandori:1993aa},
for the Moran process and a wide range of related birth death processes under weak selection \cite{nowak:2004pw,antal:2009th} 
and for some special processes for any intensity of selection \cite{antal:2009th}. 
However, Fudenberg et al. \cite{fudenberg:2006fu} obtain a slightly different result for an alternative variant of the Moran process under non-weak selection.
For large populations, the condition above reduces to risk dominance of $A$, $a_1 + a_0 > b_1 + b_0 $.
\item[(3)] {Comparison to neutrality}. 
For weak selection, the fixation probability of strategy $A$ in a finite population is larger than neutral ($1/N$)
if
$(2 N-1) a_0 + (N-2) a_1
>
(2 N-4) b_0 + (N+1) b_1$.
For a large $N$, this means that $A$ has a higher fitness than $B$ at frequency $1/3$,
termed as the one-third law \cite{nowak:2004aa,ohtsuki:2007aa,bomze:2008lr}.
The 1/3-law holds under weak selection for any process within the domain of Kingman's coalescence \cite{lessard:2007aa}.
\end{itemize}

Often, interactions are not between two players, but between whole groups of players. 
Quorum sensing,  public transportation systems or climate preservation represent examples for systems in which large groups of agents interact simultaneously.
Starting with the seminal work of Gordon and Hardin on the tragedy of the commons \cite{gordon:1954aa,hardin:1968mm},
such multiplayer games have been analyzed in the context of the evolution of cooperation \cite{hauert:1997mm,kollock:1998aa,rockenbach:2006aa,milinski:2008lr}, 
but general multiplayer interactions have received less attention, 
see however \cite{broom:1997aa,hauert:2006fd,pacheco:2009aa,kurokawa:2009aa,souza:2009aa}.

\begin{figure}
\begin{center}
\includegraphics[width=0.8 \linewidth]{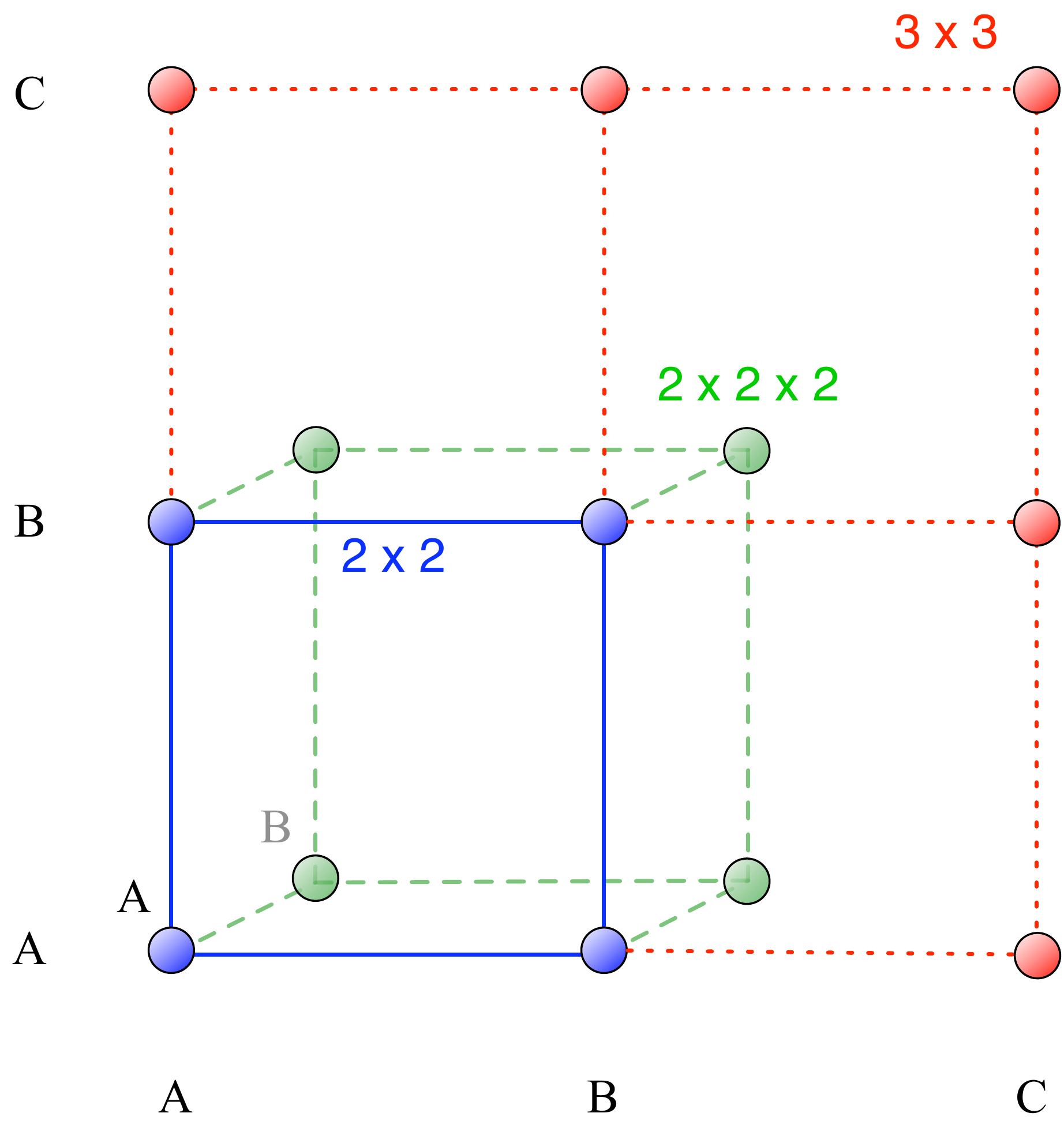}
\caption{
\label{fig:cube}
For $2 \times 2$ games, the payoff matrix has $4$ entries. 
If we increase the number of strategies, the payoff matrix grows in size. 
For example, the payoff matrix of a $3 \times 3$ game has $9$ entries. 
If we increase the number of players, the payoff matrix becomes higher dimensional.
For example, two strategy games with three players are described by $2 \times 2 \times 2$ payoff structures with 8 entries. 
In general, a $d$ player game with $n$ strategies is decribed by $n^d$ payoff values.
}
\end{center}
\end{figure}

We again assume there to be two strategies $A$ and $B$.
We can also maintain the same definition of the payoffs as $\alpha_i$.
As there are $d-1$ other individuals, excluding the focal player, $i$ can range from $0$ to $d-1$. 
We can depict the payoffs $\alpha_{i}$ in the form
\begin{align}\label{eq:Pmatrix}
\begin{array}{c cc ccc c c}
\hline\hline
\\
\rm{Opposing} \ $A$\ \rm{players} & d-1	&	& d-2	 &  \ldots 	& k &\ldots	& 0	\\
\hline	
\\
A 	& a_{d-1} &	& a_{d-2}  	&	 \ldots & a_{k} &\ldots	& a_0
 \\
 B 	&  b_{d-1} &	& b_{d-2}   	& \ldots 	& b_{k} &\ldots	& b_0	 \\
 \hline\hline
\end{array}
\end{align}
However, for multiplayer games an additional complication arises.
Consider a three player game ($d=3$).
Let the focal player be playing $A$.
As $d=3$ there are $d-1=2$ other players.
If one of them is of type $A$ and the other one of type $B$, there can be the combinations $AAB$ or $ABA$.
Do these two structures give the same payoffs? 
Or, in a more general sense, does the order of players matter?
If order does matter, the payoffs are in a $d$-dimensional discrete space as illustrated by Fig.\ \ref{fig:cube}.

There are numerous examples where the order of the players is very important.
In a game of soccer, it is necessary to have a player specialized as the goal keeper in the team. 
But it is also important that the goal keeper is at the goal and not acting as a centre-forward.
A biological example has been studied by Stander in the Etosha National Park \cite{stander:1992aa}. The lionesses hunt in packs and employ the flush and ambush technique. Some lie in ambush while others flush out the prey from the flanks and drive them towards the ones waiting in ambush. This technique needs more than two players to be successful. Some lionesses always display a particular position to be a preferred one (right flank, left flank or ambush). 
The success rate is higher if the lionesses are in their preferred positions.
Thus, the ordering of players matters here.

To address situations in which the order of player matters, we have to make use of a tensor notation for writing down the payoffs which offers the flexibility to include higher dimensions
of the payoff matrix.
Consider a tensor $\beta$ with $d$ indices defined as follows $ \beta_{i_0,i_1,i_2,i_3,....i_{d-1}}$, where the first index denotes the focal player's strategy.
Each of the indices represents the strategy of the player in the position denoted by its subscript.
The index $i$ can represent any of the $n$ strategies. Thus the total number of entries will be $n^d$.
This structure is the multiplayer equivalent to a payoff matrix, see \cite{broom:1997aa} and Fig.\ \ref{fig:cube}. 
Consider for example a game with three players and two strategies ($A$ and $B$). 
If the order of players matters, then the payoff values for strategy $A$ are represented by $\beta_{A A A}, \beta_{A A B}, \beta_{A B A}$ and $\beta_{A B B}$.
This increase in complexity is handled by the tensor notation but not reflected in the tabular notation Eq.\ \eqref{eq:Pmatrix}.
But as long as interaction groups are formed at random, we can transform the payoffs such that they can be written in the form of Eq.\ \eqref{eq:Pmatrix}, see Supporting Text. 
In this case, the payoffs are weighted by their occurrence to calculate the average payoffs. For example in our three player games, $a_1$ has to be counted twice (corresponding to $\beta_{AAB}$ and $\beta_{ABA}$).
If we would consider evolutionary games in structured populations instead of random interaction group formation, then the argument breaks down and the tensor notation cannot be reduced. 

In case of $d$ player games with two strategies we can then write the average payoff $\pi_A$ obtained by strategy $A$ in an infinite population as $\pi_A  =\sum_{k=0}^{d-1}\binom{d-1}{k} x^k(1-x)^{d-1-k} a_k $, where $x$ is the fraction of $A$ players.
An equivalent equation holds for the average payoff $\pi_B$ of strategy $B$.
The replicator equation of a 2-player game is given by \cite{hofbauer:1998mm}

\begin{align}
\label{eq:replicator}
\dot{x}=x(1-x)(\pi_A-\pi_B).
\end{align}
Obviously, there are two trivial fixed points when the whole population consists of $A$ ($x=1$) or of $B$ ($x=0$).
In $d$ player games, both $\pi_A$ and $\pi_B$ can be polynomials of maximum degree $d-1$, see Supporting Text. 
This implies that the replicator equation can have up to $d-1$ interior fixed points. 
In the two strategy case, these points can be either stable or unstable.
The maximum number of stable interior fixed points possible are $d/2$ for even $d$ and $(d-1)/2$ for odd $d$, see also \cite{hauert:2006fd} or \cite{broom:1997aa}, where it is shown that all these scenarios are also attainable.
For $d=2$, $\pi_A$ and $\pi_B$ are polynomials of degree $1$, hence there can be at most one internal equilibrium,
which is either unstable (coordination games) or stable (coexistence games).
For $d=3$, there can also be a second interior fixed point. 
If one of them is stable, the other one must be unstable.
This can lead to a situation in which $A$ is advantageous when rare (the
trivial fixed point $x=0$ is unstable), becomes disadvantageous
at intermediate frequencies, but advantageous again for high
frequencies, as in multiplayer stag hunts \cite{pacheco:2009aa}.

For a $d$ player game to have $d-1$ interior
fixed points, the quantities $a_k - b_k$ and $a_{k+1} - b_{k+1}$ 
must have different signs for all $k$. 
However, this condition is necessary (because the direction of selection can only change $d-1$ times 
if the payoff difference $a_k-b_k$ changes sign $d-1$ times), but not sufficient, see Supporting Text.
Pacheco et al. have 
studied public goods games in which a threshold frequency of cooperators is necessary for producing any public good  \cite{pacheco:2009aa,souza:2009aa}.
The payoff difference changes sign twice at this threshold value and hence there can be at most two internal equilibria.

A $d$ player game has a single internal equilibrium if  $a_k-b_k$ has a different sign than $a_{k+1}-b_{k+1}$ for a single value of $k$: In this case, $A$ individuals are disadvantageous at low frequency and advantageous at high frequency (or vice versa). If $a_k-b_k$ changes sign only once, then the direction of selection can change at most once. 
Thus, this condition is sufficient in infinite populations.

Now we deviate from the replicator dynamics,
where the average payoff of a strategy is equated to 
reproductive fitness,
and turn our attention to finite populations.
In this case, the sampling for $\pi_A$ and $\pi_B$ is no longer binomial, but hypergeometric, see Supporting Text.
In finite populations, the intensity of selection measures how important the
payoff from the game is for the reproductive fitness.
We take fitness as an exponential function of the payoff,
$f_A = \exp(+w  \pi_A)$
for $A$ players
and 
$f_B = \exp(+w  \pi_B)$
for $B$ players \cite{traulsen:2008aa}.
If $w \gg 1$, selection is strong and the average payoffs dictate the outcome of the game, whereas if $w\ll1$ then selection is weak and the payoffs have only marginal effect on the game.
This choice of fitness recovers the results of the usual Moran process introduced by Nowak et al.\
\cite{nowak:2004pw} 
and simplifies the analytical calculations significantly under strong selection \cite{traulsen:2008aa}. 
However, for non-weak selection other payoff to fitness mappings lead to slightly different results \cite{fudenberg:2006fu}.
We employ the Moran process to model the game, 
but our results hold for any birth-death process in which the ratio of transition probabilities can be approximated under weak selection by a 
term linear in the payoff difference in addition to the neutral result. 
In the Moran process, an individual is selected for reproduction at random, but proportional to its fitness. The individual produces identical offspring.  Another individual is chosen at random for death. 
With this approach we can address the basic properties of $d$ player games with $2$ strategies generalizing quantities from $2 \times 2$ games.

Does $A$ replace $B$ with a higher probability than vice versa?
Comparing 
the fixation probabilities of a single $A$ or $B$ individual,  $\rho_A$ and $\rho_B$, we find that $\rho_A > \rho_B$ 
is equivalent to
\begin{align}
\sum_{k=0}^{d-1}(N a_k - a_{d-1}) >\sum_{k=0}^{d-1} (N b_k - b_0),
\label{eq:cond2finite}
 \end{align}
 see Supporting Text.
 For $d=2$, we recover the risk dominance from above. 
 For large $N$, the condition reduces to \cite{kurokawa:2009aa} 
 \begin{align}
 \sum_{k=0}^{d-1} a_k 
 >
\sum_{k=0}^{d-1} b_k. 
\label{eq:cond2}
 \end{align}
These two conditions are valid for any intensity of selection in our variant of the Moran process.

The one third law for 2-player games is not valid for higher number of players, see Supporting Text.
Instead, the condition we obtain for the payoff entries is not directly related to the internal equilibrium points 
(as opposed to the two player case, which makes the one third law special).
For weak selection, we show in the Supporting text that $\rho_A > 1/N$ is equivalent to
\begin{align}
\label{eq:cond1finite}
\sum_{k=0}^{d-1} \left[ N (d \!- \! k) -k-1\right] a_k
>
\sum_{k=0}^{d-1}\left[ (N+1) (d \! - \! k) b_k - (d+1) b_0\right].
\end{align}
For large population size this reduces to \cite{kurokawa:2009aa}
\begin{align}
\label{eq:cond1}
 \sum_{k=0}^{d-1} (d-k) a_k  >  \sum_{k=0}^{d-1} (d-k) b_k,
\end{align}
which is the one-third law from above for $d=2$.
Inequality \eqref{eq:cond1} means that the initial phase of invasion is of most importance:
The factor $d-k$ decreases linearly with $k$ and the payoff values with small indices $k$ are more important than the payoff values with larger indices. Thus, the payoffs relevant for small mutant frequencies determine whether the condition is fulfilled.
In other words, the initial invasion is crucial to obtain a fixation probability larger than $1/N$. 

In general, the conditions \eqref{eq:cond2} and \eqref {eq:cond1} are independent of each other.
When Eq. \eqref{eq:cond2} is satisfied and Eq. \eqref{eq:cond1} is not satisfied then both fixation probabilities are less than neutral ($1/N$).
But when Eq. \eqref{eq:cond2} is not satisfied and Eq. \eqref{eq:cond1} is satisfied then both $\rho_{A}$ and $\rho_{B}$ are larger than neutral ($1/N$).
This scenario is impossible for two player games.

Let us now turn to multiplayer games with multiple strategies.
As illustrated 
in Fig.\ \ref{fig:cube}, the payoff matrix of a two player game increases in size when more strategies are added.
If more players are added, the dimensionality increases.
Now we address the evolutionary dynamics of such games. 
Again we assume that interaction groups are formed at random, such that 
only the number of players with a certain strategy -- but not their arrangement -- matters. 
The replicator dynamics of a $d$ player game with $n$ possible strategies can be written as a system of $n-1$ differential equations:
\begin{align}
\dot{x}_j= x_j (\pi_{j} - \langle\pi\rangle)
\end{align}
where $x_j$ is the frequency of strategy $j$, $\pi_j$ is the fitness of the strategy $j$ and $\langle\pi\rangle = \sum_{j=1}^{n}{x_j \pi_j}$ is the average fitness.
The evolution of this system can be studied on a simplex with $n$ vertices, $S_n$.
The simplex $S_n$ is defined by the set of all the frequencies which follow the normalisation $\sum_{j=1}^{n}{x_j} = 1$.
The fixed points of this system are given by the combination of frequencies of the strategies which satisfy,
$\pi_{1}=\cdots=\pi_{n}$.
The vertices of the simplex where $x_j$ is either equal to $1$ or $0$ are trivial fixed points.
In addition, there can be e.g.\ fixed points on the edges or the faces of the simplex. 
We speak of fixed points in the interior of the simplex when all payoffs are identical at a point where all frequencies are nonzero, 
$x_j>0$ for all $j$.
The internal equilibria are of special interest, 
because they may represent points of stable biodiversity. 
For example, 
three strains of \textit{Escherichia coli} competing for resources have been studied \cite{kerr:2002xg,czaran:2002ya}.
$K$ is a killer strain which produces a toxin harmful to $S$,
$R$ does not produce toxin, but is resistant to the toxin of $K$. 
The sensitive strain $S$ is affected by the toxin of $K$.
These three strains are engaged in a kind of rock-paper-scissors game.
$K$ kills $S$. 
$S$ reproduces faster than $R$, not paying the cost for resistance.
$R$ is superior to $K$ being immune to its toxin.
The precise nature of interactions determines whether biodiversity is maintained in an unstructured population \cite{hofbauer:1998mm,claussen:2008aa}. In our context this is reflected by the existence of an  isolated internal fixed point.

\begin{figure}
\begin{center}
\includegraphics[width=\linewidth]{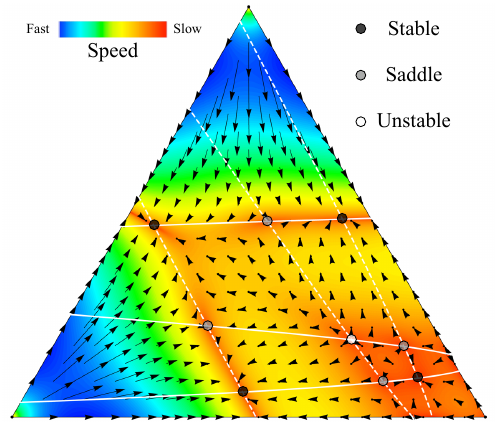}
\caption{
\label{fig:simplex}
Evolutionary dynamics in a simplex with the maximum number of internal equilibria for $d=4$ players and $n=3$ strategies as given by $(d-1)^{n-1}=9$.
On the dotted cubic curve, we have $\pi_{1} = \pi_{3}$.
On the full cubic curve, we have $\pi_{2} = \pi_{3}$.
When both lines intersect in the interior of the simplex, we have an internal equilibrium.
}
\end{center}
\end{figure}

Here, we ask the more general question whether there are internal equilibria in $d$ player games with $n$ strategies. 
If so, then how many internal equilibria are possible?
It has been shown that 
for a two player game with any number of strategies $n$ there can be at most one isolated internal equilibrium \cite{hofbauer:1998mm,bishop:1976aa}.
In the Supporting Text, we demonstrate that the maximum number of internal equilibria in $d$ players with $n$ strategies is
\begin{align}
\label{eq:maxeq}
(d-1)^{n-1}.
\end{align}
The maximum possible number of internal equilibria increases as a polynomial in the number of players, 
but exponentially in the number of strategies.
For example for $d=4$ and $n=3$ the maximum number of internal equilibria is 9, see Fig.\ \ref{fig:simplex}.

Note that for $d=2$ we recover the well known unique equilibrium.
For $n=2$, we recover the maximum of $d-1$ internal equilibria, see above. Of course, not all of these equilibria are stable. 
Broom et al.\ have shown which patterns of stability are attainable for general 3-player 3-strategy games \cite{broom:1997aa}.

This illustrates that many different states of biodiversity are possible in multiplayer games, whereas in 
two player games, only a single one is possible.
This is a crucial point when one attempts to address the question of biodiversity with evolutionary game theory.
In the previous example the studies dealing with \textit{E. coli} consider the system as a $d=2$ player game with three strategies. 
Do we really know that $d=2$?
If strains are to be engineered to stably coexist, then multiple interactions ($d>2$) would open up the possibility of multiple internal fixed points instead of the single one for $d=2$.

\begin{figure}
\begin{center}
\includegraphics[width=\linewidth]{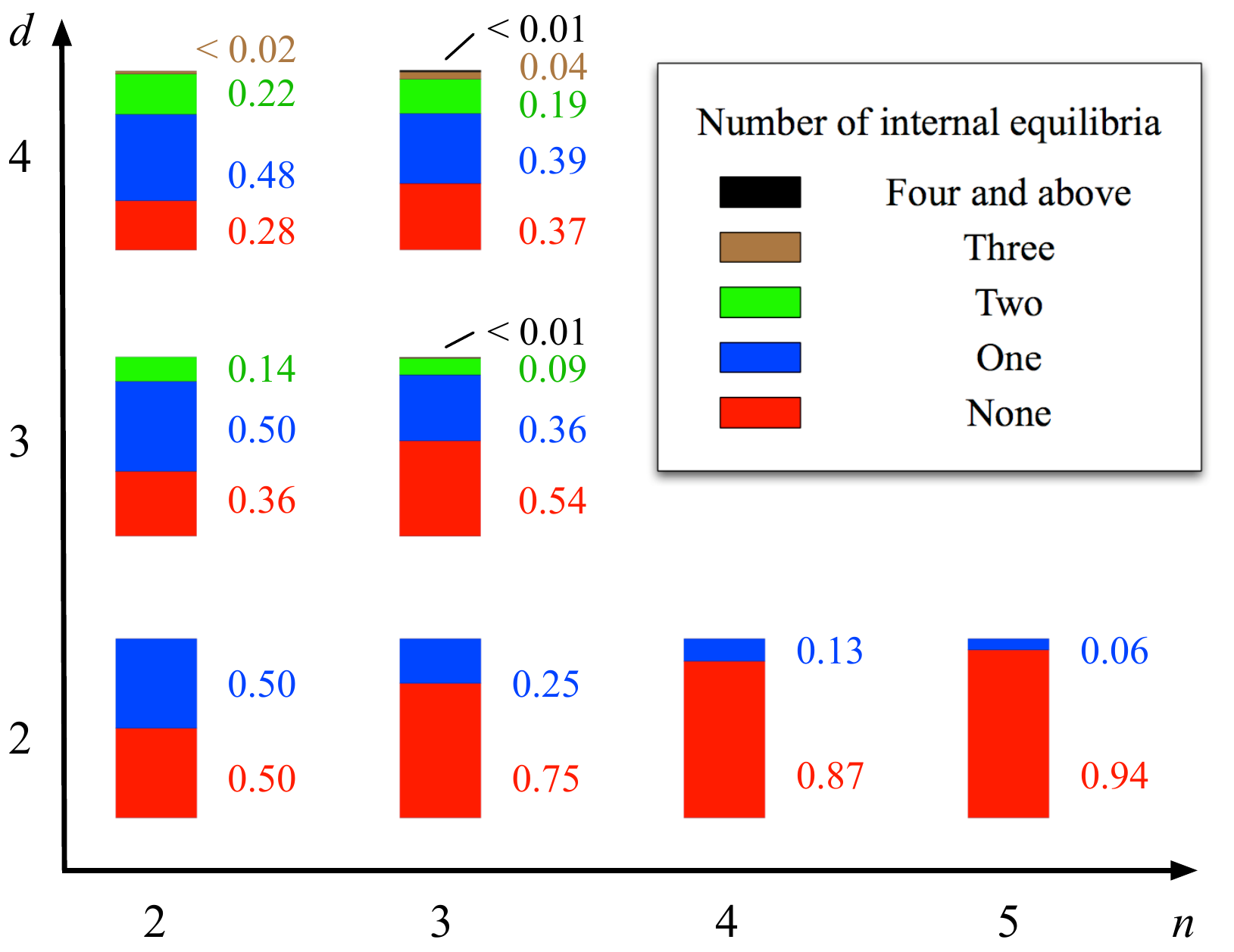}
\caption{
\label{fig:barcharts}
The probabilities of observing the different number of internal equilibria, $0$ to $(d-1)^{n-1}$ as the system gets more complex in the number of strategies $n$ and the number of players $d$. 
Random games are generated by choosing the payoff entries
$a_k, b_k, \ldots$
 from a uniform distribution.
If we consider that the order does matter and generate the random games based on the entries of a payoff structure with $n^d$ entries, then the probability of observing a particular number of equilibria is only slightly lower
(averages over $10^{8}$ different games with uniformly chosen payoff entries $a_k, b_k, \ldots$).
}
\end{center}
\end{figure}

If we choose a game at random, what is the probability that the game has a certain number of internal equilibria?
To this end, we take the following approach:  
We generate many random payoff structures in which all payoff entries are uniformly distributed random numbers \cite{huang:2010aa}.
For each payoff structure, we compute the number of internal equilibria.
It turns out that games with many internal equilibria are the exception rather than the rule.
For example, the probability to see $2$ or more internal equilibria in a game with $4$ players and $3$ strategies is $\approx 24 \%$. 
The probability that a randomly chosen game has the
maximum possible number of equilibria 
decreases with increasing number of players and number of strategies,
see Fig.\ \ref{fig:barcharts}.

Also the probability of having a single equilibrium reduces. Instead we obtain several internal equilibria in the case
of more than two players. 
For two player games, the probability to see an internal equilibrium at all decreases roughly exponentially with the number of strategies. 
This poses an additional difficulty in coordinating in multiplayer games, because several different solutions may be possible that look quit similar at first sight.

\section{Discussion}

Multiplayer games with multiple strategies is what we find all around. 
We interact with innumerable people at the same time, directly or indirectly. 
Some interactions may be pairwise, but others are not.
In real life, we may typically be engaged in many person games instead of a disjoined collection of two person games \cite{hamilton:1975aa}.
The evolution and maintenance of cooperation, problems pertaining from group hunting to deteriorating climate, all are fields for a multiple number of players \cite{stander:1992aa,levin:2009aa,milinski:2008lr,broom:2003aa}.
They can have different interests and hence use different strategies.
There are other cases like the maintenance of biodiversity where 
multiplayer interactions may lead to a much richer spectrum for biodiversity than the commonly 
analyzed two player interactions.
The presence of multiple stable states
also contributes to the intricate dynamics observed in maintenance of biodiversity \cite{levin:2000aa}.
Multiplayer games may help to improve our understanding of such systems.
The problem of handling multiple equilibria is not just limited to biological games but it also appears in economics \cite{kreps:1990bo,damme:1994aa}.
Many insights can be obtained by studying two player games,  
but it blurs the complexity of multiplayer interactions.
Here, we have derived some basic rules which apply to multiplayer games with two strategies for finite as well as infinite populations and discussed the number of internal equilibria in $d$ player game with $n$ strategies, which determine how the dynamics proceeds.

This theory can be applied to all kinds of games with any number of players and strategies 
and can thus be easily applied to public goods games, multiplayer stag hunts or multiplayer snowdrift games.
We believe that this opens up new avenues where we can get analytical description of situations which are thought to be very complex 
and further discussions on these issues will prove to be fruitful due to the intrinsic importance of multiplayer interactions.
We conclude this approach by quoting Hamilton again, ``A healthy society \textit{should} feel sea-sick when confronted with the endless internal instabilities of the `solutions', `coalition sets', etc., which the theory of many-person games has had to describe." \cite{hamilton:1975aa}.

\begin{acknowledgments}
We thank the anonymous referees for their helpful comments.
C.S.G. and A.T. acknowledge support by the Emmy-Noether program of the Deutsche Forschungsgemeinschaft
and the DAAD (project 0813008).
\end{acknowledgments}


\begin{thebibliography}{10}

\bibitem{maynard-smith:1973to}
Maynard~Smith, J, Price, GR
\newblock (1973) The logic of animal conflict.
\newblock {\em Nature} 246:15--18.

\bibitem{maynard-smith:1982to}
Maynard~Smith, J
\newblock (1982) {\em Evolution and the Theory of Games}
\newblock (Cambridge University Press, Cambridge).

\bibitem{nowak:2006bo}
Nowak, MA
\newblock (2006) {\em Evolutionary Dynamics}
\newblock (Harvard University Press, Cambridge, MA).

\bibitem{nowak:2006pw}
Nowak, MA
\newblock (2006) Five rules for the evolution of cooperation.
\newblock {\em Science} 314:1560--1563.

\bibitem{taylor:2007bb}
Taylor, C, Nowak, MA
\newblock (2007) Transforming the dilemma.
\newblock {\em Evolution} 61:2281--2292.

\bibitem{ostrom:1990bo}
Ostrom, E
\newblock (1990) {\em Governing the Commons: The Evolution of Institutions for
  Collective Action}
\newblock (Cambridge Univ. Press).

\bibitem{hauert:2002te}
Hauert, C, De~Monte, S, Hofbauer, J, Sigmund, K
\newblock (2002) Volunteering as red queen mechanism for cooperation in public
  goods games.
\newblock {\em Science} 296:1129--1132.

\bibitem{hamilton:1975aa}
Hamilton, WD
\newblock (1975) in {\em Biosocial Anthropology}, ed{} Fox, R
\newblock (Wiley, New York), pp 133--155.

\bibitem{broom:1997aa}
Broom, M, Cannings, C, Vickers, G
\newblock (1997) Multi-player matrix games.
\newblock {\em Bull. Math. Biol.} 59:931--952.

\bibitem{hauert:2006fd}
Hauert, C, Michor, F, Nowak, MA, Doebeli, M
\newblock (2006) Synergy and discounting of cooperation in social dilemmas.
\newblock {\em J. Theor. Biol.} 239:195--202.

\bibitem{pacheco:2009aa}
Pacheco, JM, Santos, FC, Souza, MO, Skyrms, B
\newblock (2009) Evolutionary dynamics of collective action in n-person stag
  hunt dilemmas.
\newblock {\em Proc. R. Soc. B} 276:315--321.

\bibitem{souza:2009aa}
Souza, MO, Pacheco, JM, Santos, FC
\newblock (2009) Evolution of cooperation under n-person snowdrift games.
\newblock {\em J. Theor. Biol.} 260:581--588.

\bibitem{kurokawa:2009aa}
Kurokawa, S, Ihara, Y
\newblock (2009) Emergence of cooperation in public goods games.
\newblock {\em Proc. R. Soc. B} 276:1379--1384.

\bibitem{veelen:2009ma}
van Veelen, M
\newblock (2009) Group selection, kin selection, altruism and cooperation: when
  inclusive fitness is right and when it can be wrong.
\newblock {\em J. Theor. Biol.} 259:589--600.

\bibitem{nowak:2004pw}
Nowak, MA, Sasaki, A, Taylor, C, Fudenberg, D
\newblock (2004) Emergence of cooperation and evolutionary stability in finite
  populations.
\newblock {\em Nature} 428:646--650.

\bibitem{kandori:1993aa}
Kandori, M, Mailath, GJ, Rob, R
\newblock (1993) Learning, mutation, and long run equilibria in games.
\newblock {\em Econometrica} 61:29--56.

\bibitem{antal:2009th}
Antal, T, Nowak, MA, Traulsen, A
\newblock (2009) Strategy abundance in 2x2 games for arbitrary mutation rates.
\newblock {\em J. Theor. Biol.} 257:340--344.

\bibitem{fudenberg:2006fu}
Fudenberg, D, Nowak, MA, Taylor, C, Imhof, L
\newblock (2006) Evolutionary game dynamics in finite populations with strong
  selection and weak mutation.
\newblock {\em Theor. Pop. Biol.} 70:352--363.

\bibitem{nowak:2004aa}
Nowak, MA, Sigmund, K
\newblock (2004) Evolutionary dynamics of biological games.
\newblock {\em Science} 303:793--799.

\bibitem{ohtsuki:2007aa}
Ohtsuki, H, Bordalo, P, Nowak, MA
\newblock (2007) The one-third law of evolutionary dynamics.
\newblock {\em J. Theor. Biol.} 249:289--295.

\bibitem{bomze:2008lr}
Bomze, I, Pawlowitsch, C
\newblock (2008) One-third rules with equality: Second-order evolutionary
  stability conditions in finite populations.
\newblock {\em J. Theor. Biol.} 254:616--620.

\bibitem{lessard:2007aa}
Lessard, S, Ladret, V
\newblock (2007) The probability of fixation of a single mutant in an
  exchangeable selection model.
\newblock {\em J. Math. Biol.} 54:721--744.

\bibitem{gordon:1954aa}
Gordon, HS
\newblock (1954) The economic theory of a common-property resource: The
  fishery.
\newblock {\em The Journal of Political Economy} 62:124--142.

\bibitem{hardin:1968mm}
Hardin, G
\newblock (1968) The tragedy of the commons.
\newblock {\em Science} 162:1243--1248.

\bibitem{hauert:1997mm}
Hauert, C, Schuster, HG
\newblock (1997) Effects of increasing the number of players and memory size in
  the iterated {P}risoner's {D}ilemma: {A} numerical approach.
\newblock {\em Proc. R. Soc. Lond. B} 264:513--519.

\bibitem{kollock:1998aa}
Kollock, P
\newblock (1998) Social dilemmas: The anatomy of cooperation.
\newblock {\em Annual Review of Sociology} 24:183--214.

\bibitem{rockenbach:2006aa}
Rockenbach, B, Milinski, M
\newblock (2006) The efficient interaction of indirect reciprocity and costly
  punishment.
\newblock {\em Nature} 444:718--723.

\bibitem{milinski:2008lr}
Milinski, M, Sommerfeld, RD, Krambeck, HJ, Reed, FA, Marotzke, J
\newblock (2008) The collective-risk social dilemma and the prevention of
  simulated dangerous climate change.
\newblock {\em Proc. Natl. Acad. Sci. USA} 105:2291--2294.

\bibitem{stander:1992aa}
Stander, PE
\newblock (1992) Cooperative hunting in lions: the role of the individual.
\newblock {\em Behavioral Ecology and Sociobiology} 29:445--454.

\bibitem{hofbauer:1998mm}
Hofbauer, J, Sigmund, K
\newblock (1998) {\em Evolutionary Games and Population Dynamics}
\newblock (Cambridge University Press, Cambridge).

\bibitem{traulsen:2008aa}
Traulsen, A, Shoresh, N, Nowak, MA
\newblock (2008) Analytical results for individual and group selection of any
  intensity.
\newblock {\em Bull. Math. Biol.} 70:1410--1424.

\bibitem{kerr:2002xg}
Kerr, B, Riley, MA, Feldman, MW, Bohannan, BJM
\newblock (2002) Local dispersal promotes biodiversity in a real-life game of
  rock-paper-scissors.
\newblock {\em Nature} 418:171--174.

\bibitem{czaran:2002ya}
Czaran, TL, Hoekstra, RF, Pagie, L
\newblock (2002) Chemical warfare between microbes promotes biodiversity.
\newblock {\em Proc. Natl. Acad. Sci. USA} 99:786--790.

\bibitem{claussen:2008aa}
Claussen, JC, Traulsen, A
\newblock (2008) Cyclic dominance and biodiversity in well-mixed populations.
\newblock {\em Phys. Rev. Lett.} 100:058104.

\bibitem{bishop:1976aa}
Bishop, DT, Cannings, C
\newblock (1976) Models of animal conflict.
\newblock {\em Advances in Applied Probability} 8:616--621.

\bibitem{huang:2010aa}
Huang, W, Traulsen, A
\newblock (2010) Fixation probabilities of random mutants under frequency
  dependent selection.
\newblock {\em J. Theor. Biol.} in press.

\bibitem{levin:2009aa}
Levin, SA, ed
\newblock (2009) {\em Games, Groups and the Global Good}, Springer Series in
  Game Theory
\newblock (Springer).

\bibitem{broom:2003aa}
Broom, M
\newblock (2003) The use of multiplayer game theory in the modeling of
  biological populations.
\newblock {\em Comments on Theoretical Biology} 8:103--123.

\bibitem{levin:2000aa}
Levin, SA
\newblock (2000) Multiple scales and the maintenance of biodiversity.
\newblock {\em Ecosystems} 3:498--506.

\bibitem{kreps:1990bo}
Kreps, DM
\newblock (1990) {\em Game Theory and Economic Modelling (Clarendon Lectures in
  Economics)}
\newblock (Oxford University Press, USA).

\bibitem{damme:1994aa}
Damme, EV
\newblock (1994) Evolutionary game theory.
\newblock {\em European Economic Review} 38:847--858.

\end{thebibliography}
\end{document}


\title{\sffamily Supporting Information:\\ Evolutionary games in the multiverse}

\author{Chaitanya S. Gokhale}
\address{Emmy-Noether Group for Evolutionary Dynamics,
Department of Evolutionary Ecology,
Max-Planck-Institute for Evolutionary Biology, August-Thienemann-Stra{\ss}e 2, 24306 Pl\"{o}n, Germany}
\author{Arne Traulsen}
\email{traulsen@evolbio.mpg.de}
\address{Emmy-Noether Group for Evolutionary Dynamics,
Department of Evolutionary Ecology,
Max-Planck-Institute for Evolutionary Biology, August-Thienemann-Stra{\ss}e 2, 24306 Pl\"{o}n, Germany}

\sffamily

\maketitle


\section{Multiple players with two strategies}
\subsection{Infinite populations}
\label{infinitepop}
We first address the replicator dynamics of multiplayer games with two strategies. 
If an $A$-player interacts with $k$ other $A$-players, it obtains the payoff $a_{k}$.
If a $B$-player interacts with $k$ $A$-players, it obtains the payoff $b_{k}$.
In an infinitely large population in which the fraction of $A$-players is $x$, the
probability that an $A$-player interacts with $k$ other $A$-players  
is 
\begin{align}
\binom{d-1}{k} x^k(1-x)^{d-1-k}. 
\end{align}
Here, $\binom{d-1}{k}$ is the number of possibilities
to arrange the players. Thus, the average payoffs of $A$ and $B$ are given by
\begin{align}
\label{payoff}
\pi_A & =\sum_{k=0}^{d-1}\binom{d-1}{k} x^k(1-x)^{d-1-k} a_k  \\ \nonumber
\pi_B & =\sum_{k=0}^{d-1}\binom{d-1}{k} x^k(1-x)^{d-1-k} b_k. 
\end{align}
These average payoffs are subject to the condition that the order of the players does not matter.
For example, in a $d=3$ game let the player in first position play $A$. Then, the remaining two players can play a combination of $A$ and $B$. The possible combinations are $AAB$ and $ABA$. By writing the payoffs in the above mentioned manner we assume that such combinations have the same payoffs. 

If the order of players does matter, then 
the payoff values are given by $ \beta_{i_0,i_1,i_2,i_3,....i_{d-1}}$. 
Here, $i_{0}$ is the strategy of the focal player.
The $i_{p}$ are the strategies of the type in position $p$.
For random matching of players, we can map the $ \beta_{i_0,i_1,i_2,i_3,....i_{d-1}}$ to modified payoffs $\tilde{a}_k$ and $\tilde{b}_k$ without changing the average payoffs of the strategies.
As an example, for $d=4$, we have the modified payoffs $\tilde{a}_k$ and $\tilde{b}_k$ as,
\begin{align}
\tilde{a}_0 &=\beta_{A,B,B,B} &\tilde{b}_0&= \beta_{B,B,B,B} \\ \nonumber
\tilde{a}_1 &= \frac{\beta_{A,A,B,B} +\beta_{A,B,A,B}+\beta_{A,B,B,A}}{3} &\tilde{b}_1 &= \frac{\beta_{B,A,B,B} +\beta_{B,B,A,B}+\beta_{B,B,B,A}}{3} \\ \nonumber
\tilde{a}_2 &= \frac{\beta_{A,A,A,B} +\beta_{A,A,B,A}+\beta_{A,B,A,A}}{3} &\tilde{b}_2 &= \frac{\beta_{B,A,A,B} +\beta_{B,A,B,A}+\beta_{B,B,A,A}}{3} \\ \nonumber
\tilde{a}_3 &= \beta_{A,A,A,A} &\tilde{b}_3 &= \beta_{B,A,A,A} \\ \nonumber
\end{align}

We just need to substitute the above payoffs in place of $a_k$ and $b_k$ in Eqs. \eqref{payoff} to take into account the effect of the arrangement of players.
For any number of players such a generalization can be easily obtained.
Thus, the evolutionary dynamics under random interaction group formation remains unaffected of the fact that the order of players does matter.
When interaction groups are not formed at random, this argument will of course fail in most cases. 

The following analysis deals with $\pi_A$ and $\pi_B$ as in Eq. \eqref{payoff}, but it also holds when the order of players matters, but interaction groups are formed at random.
The replicator equation is thus given by  \cite{taylor:1978wv,hofbauer:1998mm}
\begin{equation}
\label{replicator}
\dot{x}=x(1-x)(\pi_A-\pi_B)
\end{equation}
Both $\pi_A$ and $\pi_B$ are polynomials of degree $d-1$. 
This implies that the replicator equation can have up to $d-1$ interior fixed points \cite{hauert:2006fd}. 

\subsubsection*{Maximum number of interior fixed points}

For a $d$ player game to have $d-1$ interior
fixed points, the quantities $a_k - b_k$ and $a_{k+1} - b_{k+1}$ 
must have different sign for all $k$. 
For example, in a three-player game with  
$a_0=+1$, $a_1=-\lambda$, $a_2=+1$ and
$b_0=-1$, $b_1=+\lambda$, $b_2=-1$, we have two internal equilibria at $\frac{1}{2}\left(1 \pm \sqrt{\frac{\lambda-1}{\lambda +1}}\right)$ for $\lambda>1$.
However, this condition is necessary (because the direction of selection can only change $d-1$ times 
if the payoff difference $a_k-b_k$ changes sign $d-1$ times), but not sufficient. 
For example, in the above three player game, there are no internal equilibria for $\lambda < 1 $.

\subsubsection*{Single interior fixed point}

A $d$ player game has a single internal equilibrium if  $a_k-b_k$ has a different sign than $a_{k+1}-b_{k+1}$ for a single value of $k$: In this case, A individuals are disadvantageous at low frequency and advantageous at high frequency (or vice versa). If $a_k-b_k$ changes sign only once, then the direction of selection can obviously at most change once. Thus, this condition is sufficient.

\subsection{Finite populations}
Let us now turn to the evolutionary dynamics in finite populations. 
In a population of size $N$ with $j$ individuals of type $A$, the
probability to choose a group that consists of $k$ $A$ players
and $d-1-k$ $B$ players is given by a hypergeometric distribution.
The probability that an $A$ player interacts with $k$ other $A$ players is given by
\begin{align}
H(k,d;j,N) = \frac{ \binom{j-1}{k} \binom{N-j}{d-1-k} }{\binom{N-1}{d-1}}.
\end{align}
This leads to the average payoffs
\begin{align}
\label{payofffin}
\pi_A &=\sum_{k=0}^{d-1}  \frac{ \binom{j-1}{k} \binom{N-j}{d-1-k} }{\binom{N-1}{d-1}} a_k \\ \nonumber
\pi_B & =\sum_{k=0}^{d-1}  \frac{ \binom{j}{k} \binom{N-j-1}{d-1-k} }{\binom{N-1}{d-1}} b_k.
\end{align}

We assume that strategies spread by a frequency dependent Moran process \cite{moran:1962ef,ewens:2004qe,nowak:2004aa}. 
The fitness is given by 
$f_A = \exp(+w  \pi_A)$
for $A$ players and 
$f_B = \exp(+w  \pi_B)$
for $B$ players,
where $w$ measures the intensity of selection \cite{traulsen:2008aa}. 
For $w \ll 1$, selection is weak.
For $w \gg 1$, selection is strong and only the fitter type reproduces. 
 In the Moran process, an individual is selected for reproduction at random, but proportional to its fitness. The individual produces identical offspring. 
Another individual is chosen at random for death. 
Consider $j$ individuals of type $A$ in a population of size $N$. 
The number of $A$ individuals increase with probability $T^+_j$ from $j$ to $j+1$ if an $A$ individual is selected for reproduction and a $B$ individual dies. We have
\begin{align}
\label{eq:tp1}
T_j^+  &= \frac{j f_A}{j f_A + (N-j) f_B} \frac{N-j}{N} \\
T_j^- &= \frac{(N-j) f_B}{j f_A + (N-j) f_B} \frac{j}{N}. 
\end{align}
The fixation probability of a single $A$ individual in a population of $N$ is given by \cite{nowak:2006bo},
\begin{equation}
\label{eq:phione}
\rho_A = \frac{1}{1+\sum_{m=1}^{N-1}\prod_{j=1}^{m}\frac{T_j^-}{T_j^+}}.
\end{equation}
For the ratio of transition probabilities, we have
\begin{align}
\frac{T_j^-}{T_j^+} 
= \frac{f_B}{f_A}
= e^{-w(\pi_A - \pi_B)} 
\approx 1 - w(\pi_A - \pi_B).
\end{align}
The approximation is valid for weak selection, $w \ll 1$. 
Note that this is the only approximation we make, such that our result is valid for any birth-death process with
\begin{equation}
\frac{T_j^-}{T_j^+} \approx 1 - w(\pi_A - \pi_B).
\label{wsapprox}
\end{equation}
For weak selection, the product in the fixation probabilities can be approximated by a
sum, which leads to 
\begin{equation}
\label{eq:phioneapprox}
\rho_A  \approx  \frac{1}{N}+\frac{w}{N^2}\underbrace{\sum_{m=1}^{N-1}\sum_{j=1}^{m}\left( \pi_A - \pi_B \right)}_{\Gamma}.
\end{equation}
In Appendix \ref{appcond1} we show that,
\begin{align}
\Gamma=\frac{1}{d (d+1)} \left[N^{2} \left(\sum_{k=0}^{d-1} (d-k) (a_k-b_k)\right)-N \left(\sum_{k=0}^{d-1} (k+1) a_k + \sum_{k=1}^{d-1} (d-k) b_k - d^{2} b_{0}\right)\right].
\end{align}
As seen from Eq. \eqref{eq:phioneapprox}, a strategy is favored by selection, i.e. it has a fixation probability larger than $1/N$ if 
$\Gamma>0$.
For any $N$, $\Gamma>0$ can be represented by,
\begin{align}
\label{finiteNcond1}
\sum_{k=0}^{d-1} \left[ N (d-k) -k-1\right] a_k > \sum_{k=0}^{d-1}\left[ (N+1) (d-k) b_k -(d+1) b_0\right]
\end{align}
For $d=2$, this condition reduces to the condition, $(2 N-1)a_0 + (N-2)a_1 >(2 N-4)b_0 + (N+1)b_1$, exactly  as developed by Nowak et. al. \cite{nowak:2004pw}.
For a large population size the condition can be simplified to
\begin{align}
\label{eq:cond1}
\sum_{k=0}^{d-1} (d-k) a_k  > \sum_{k=0}^{d-1} (d-k) b_k.
\end{align}
In large populations, we have $\rho_A >1/N$ if the condition \eqref{eq:cond1} is fulfilled.
In the usual case of $d=2$, the fixation probability of strategy $A$ is larger than $1/N$ if
$2 a_0+a_1 > 2 b_0+b_1$. This can be rearranged to
\begin{align}
x^{\ast} = \frac{b_0-a_0}{a_1-a_0-b_1+b_0} < \frac{1}{3}.
\end{align}
This is the $1/3$-law first derived in \cite{nowak:2004pw}: A mutant
takes over the population with probability larger than neutral if the mutant
is advantageous when it has reached a fraction of $1/3$. 
Equation \eqref{eq:cond1} represents a generalization of the $1/3$-law for general
$d$-player games. 

We can also compare the fixation probability $\rho_A$ of a single $A$ player  
to the fixation probability $\rho_B$ of a single $B$ player. It has been shown \cite{nowak:2006bo, traulsen:2008aa}
that
\begin{align}
\label{eq:ratio}
\frac{\rho_B}{\rho_A} = \prod_{j=1}^{N-1}\frac{T_j^-}{T_j^+} 
= \exp \left[ - w \underbrace{\sum_{j=1}^{N-1}(\pi_A-\pi_B)}_{\Phi} \right]
\end{align}
Note that if our previous approximation Eq.\ \eqref{wsapprox} holds, then we obtain 
$\frac{\rho_B}{\rho_A} \approx1- w \Phi$. Since we do not make any further approximations, 
our calculation remains valid for any birth-death process fulfilling Eq.\ \eqref{wsapprox} under weak selection.
As shown in Appendix  \ref{appcond2},
\begin{align}
\Phi &=\frac{N}{d} \sum_{k=0}^{d-1}(a_k-b_k) + b_0-a_{d-1}
\end{align}
From Eq. \eqref{eq:ratio}, it is clear that $\rho_A > \rho_B$ if $\Phi>0$.
This is equivalent to the condition
\begin{align}
\sum_{k=0}^{d-1}(N a_k - a_{d-1}) >\sum_{k=0}^{d-1} (N b_k - b_0)
\label{grgrka}
\end{align}
Note that this condition is valid for any intensity of selection for the process we use.
For weak selection, it is valid for all processes with $\frac{T_j^-}{T_j^+} 
\approx 1 - w (\pi_A-\pi_B)$.
For $d=2$, Eq. \eqref{grgrka} reduces to $(N-2)(a_1-b_0)>N(b_1-a_0)$ which is the risk dominance condition developed in \cite{kandori:1993aa} for finite population size, see also \cite{antal:2009th} for the generality of this finding.
For a large population the condition can be further simplified,
\begin{align}
 \sum_{k=0}^{d-1} a_k 
 >
\sum_{k=0}^{d-1} b_k. 
\label{eq:cond2}
 \end{align}
For 2-player games, this reduces to risk dominance, $a_0 +a_1> b_0+b_1$.

We can also incorporate mutations, which will complicate the transition probabilities.
For symmetric mutation rates, $\mu_{A \to B} = \mu_{B \to A}$, 
the condition $\rho_A > \rho_B$ is equivalent to a higher
average abundance of $A$ compared to $B$ given that $\mu_{A \to B}$ and $\mu_{B \to A}$ are small. 
For $d=2$, it has recently been shown that
the abundance condition does in fact neither depend on the mutation rate nor on the
intensity of selection \cite{antal:2009th}. For $d>2$, this statement does no longer hold, which
can be seen from the high mutation limit: If the mutation rates are very high, then the system
will be driven towards the point where the two abundances are identical. The dynamics at this point, 
however, does not depend on the parameters in the same way as $\rho_A > \rho_B$ when it comes
to $d$ player games.

\section{Multiplayer games with multiple strategies}
\subsection{Infinite populations}
In the full multiverse, we have multiple players playing multiple strategies.
We are interested in the maximum number of internal equilibria of a system, which will help us understand the general features of the dynamics.
Consider a system with $d$ players with $n$ possible strategies.
Here we resort to the payoff values as given by $\beta_{i_0,i_1,i_2,i_3,....i_{d-1}}$,
 because for random group formation a system where the order of players does matter can always be reduced to a system where the order does not matter.
Here, $i_{0}$ is the strategy of the focal player.
The $i_{p}$ are the strategies of the type in position $p$.
The the average payoff of the focal player is given by
\begin{align}
\pi_{i_{0}} = \sum_{i_1 = 1}^{n} \sum_{i_2 = 1}^{n} \ldots \sum_{i_{d-1} = 1}^{n} \left( \prod_{k = i_1}^{i_{d-1}} x_k\right) \beta_{i_0,i_1,i_2,i_3,....i_{d-1}}.
\end{align}
From this it is clear that each variable $x_k$ is at most of degree $d-1$.
Also as there are $n$ strategies we have, $i_0 = (1,2, \ldots n)$, that is $n$ such multivariate polynomials.
Each multivariate polynomials is in $n-1$ variables (because of the normalization $\sum_{l = 1}^{n}{x_l } = 1$).
At the fixed points, all these polynomials will be equal. Hence if we subtract one of the polynomials (say $\pi_{n}$) from all, we have a system of $n-1$ multivariate polynomials, $\Delta \pi_{i_0}$, equal to zero (where $i_0$ goes from $1$ to $n-1$).
In each variable $x_k$, the multivariate polynomial $\Delta \pi_{i_0}$ is at most of degree $d-1$. Hence, there are at most $d-1$ roots of $\Delta \pi_{i_0}$ in $x_k$.
Since this is valid for all $n-1$ functions of $\Delta \pi_{i_0}$, there can be up to $(d-1)^{n-1}$ simultaneous roots of all $\Delta \pi_{i_0}$.
These are the interior fixed points of the replicator dynamics.
Thus, there can be at most 
 \begin{align}
 (d-1)^{n-1}
 \end{align}
 fixed points in the interior of the system. This holds for the full system but also
 for any subspace in which less strategies are available. For example, a game with
 $d=3$ players and $n=4$ strategies has up to 8 fixed points in the interior of
 the simplex $S_{4}$. 
On the faces of the simplex $S_{4}$, represented by the simplex $S_{3}$,  there can be up to 4 fixed points.

We now have an analytic method to deduce the maximum number of internal equilibria.
The question that now arises is with what probability do we see this maximum number of equilibria?
We address the problem by generating $10^8$ payoff matrices where the payoff values $a_k$, $b_k$, $\ldots$ are drawn from a uniform distribution for different configurations of $d$ and $n$.
As discussed in the main text, the probability to obtain the maximum number of internal equilibria in a game with random payoff entries reduces as the complexity increases in $d$ as well as $n$.

\subsubsection*{An example for $\boldsymbol{ d=4}$ and $\boldsymbol{ n=3}$}
In this section, we describe the parameters of Fig.\ 2 in the main text.
The number of players $d=4$ and the number of strategies is $n=3$.
The total number of payoff values is therefore $n^d$ which is $81$.
Thus for each strategy there are $27$ payoff values.
This is the number of values we have to consider when the order of player matters.
If the payoffs are the same for different arrangements then we reduce the payoff values, but we have to weight them by the number of their occurrence.
Consider the three strategies to be $A$, $B$ and $C$.
Solving the replicator equation using the average payoffs calculated from the payoffs from Table \ref{fig:simplexpayoffs}, we numerically obtain $9$ fixed points in the interior of the simplex.
At these points the frequencies of all the strategies are non-zero and the average payoff to each strategy is equal.

\begin{table}[h]
\begin{center}
\begin{tabular}{ccccccccccc}
\hline 
\begin{minipage}[c]{2cm}
\begin{center}
\vspace{.1cm}
Weight\vspace{-.1cm}
 \\ 
(Total 27)
\vspace{.2cm}
\end{center}
\end{minipage}
 & 1 & 3 & 3 & 3 & 6 & 3 & 1& 3& 3&  1 \\
 \hline
Configuration & AAA & AAB & AAC & ABB & ABC & ACC & BBB & BBC & BCC & CCC \\
\hline
A & -9.30 & 3.83 & 3.86 & -1.03 & -1.00 & -0.96 & 0.10 & 0.33 & 0.16 & 0.20 \\
B & 0.10 & -1.03 & 0.13 & 3.83 & -1.00 & 0.16 & -9.30 & 4.06 & -0.96 & 0.20\\
C & 0.00 & 0.00 & 0.00 & 0.00 & 0.00 & 0.00 & 0.00 & 0.20 & 0.00 & 0.00 \\
\hline
\end{tabular}
\caption{
\label{fig:simplexpayoffs}
The reduced payoff table for the $d=4$ and $n=3$ game in Fig.\ 2 in the main text.
In total, there would be  $n^d = 3^4 = 81$ payoff entries.
For each strategy, we would have had $27$ entries.
But when we consider that the ordering does not matter, then we just weight each configuration by the different ways of ordering, e.g. there are three orderings for $AAB$, i.e.\ $AAB$, $ABA$, and $BAA$.
In this way, we reduce the number of payoff entries from $81$ to $30$.
}
\end{center}
\end{table}

\subsection{Finite populations}

For finite populations and more than two strategies, few analytical tools are available. 
The average abundance under weak selection can be addressed using tools from coalescence theory
\cite{antal:2009aa,antal:2009sf}.

For small mutation rates, the dynamics reduces to an embedded Markov chain on the pure states of the system, see Fudenberg and Imhof  \cite{fudenberg:2006ee} for a proof.
Essentially, this means that the dynamics is governed by dynamics on the edges of the simplex $S_{n}$, where
only two strategies are present. This result can be applied in a variety of contexts \cite{imhof:2005oz,hauert:2007aa,van-segbroeck:2009mi}.

Both approaches can be adapted to $d$ player games.

\appendix

\section{Condition for the comparison of one strategy with neutrality}
\label{appcond1}
We first repeat the condition to prove,
\begin{align}
&\sum_{m=1}^{N-1}\sum_{j=1}^{m}\left( \pi_A - \pi_B \right) \nonumber \\
&=\frac{1}{d (d+1)} \left[N^{2} \left(\sum_{k=0}^{d-1} (d-k) (a_k-b_k)\right) -N \left(\sum_{k=0}^{d-1} (k+1) a_k + \sum_{k=1}^{d-1} (d-k) b_k - d^{2} b_{0}\right)\right],
\end{align}
where the payoffs are defined in Eq.\ \eqref{payofffin}.
Since all the $a_k$'s come from $\pi_A$ and all the $b_k$'s from $\pi_B$ we can solve each separately.
For $\pi_A$ we have to show that,
\begin{align}
\sum_{m=1}^{N-1}\sum_{j=1}^{m} \sum_{k=0}^{d-1} \frac{\binom{j-1}{k} \binom{N-j}{d-k-1}}{\binom{N-1}{d-1}} a_k
=  \sum_{k=0}^{d-1}    \frac{N^2(d-k)-N(k+1)}{d(d+1)} a_k.
\end{align}
Since this should hold for any choice of the $a_k$'s, we must show that 
\begin{align}
\label{eq:toprove2}
\sum_{m=1}^{N-1}\sum_{j=1}^{m}\frac{\binom{j-1}{k} \binom{N-j}{d-k-1}}{\binom{N-1}{d-1}} 
=    \frac{N^2(d-k)-N(k+1)}{d(d+1)}. 
\end{align}
We take out the factor $\binom{N-1}{d-1}^{-1}$ on the left hand side and get back to the full expression only at the end. 
We consider the quantity
\begin{align}
\sum_{m=1}^{N-1}\sum_{j=1}^{m}  \binom{j-1}{k} \binom{N-j}{d-k-1}.
\end{align}
Using the identity $\sum_{m=1}^{N-1}\sum_{j=1}^{m}=\sum_{j=1}^{N-1}\sum_{m=j}^{N-1}$, we obtain
\begin{align}
\label{eq16}
\sum_{m=1}^{N-1}\sum_{j=1}^{m}   \binom{j-1}{k} \binom{N-j}{d-k-1} 
&= \sum_{j=1}^{N-1} \sum_{m=j}^{N-1} \binom{j-1}{k} \binom{N-j}{d-k-1} \\ \nonumber
&= \sum_{j=1}^{N-1}\binom{j-1}{k} \binom{N-j}{d-k-1} (N-j) ,
\end{align}
where we performed the sum over $m$. 
Let us use the factor $N-j$ to split this expression into two sums. 
The first sum with the factor $N$ is given by
\begin{align}
\label{eq27}
\Sigma_1=
N \sum_{j=1}^{N-1}\binom{j-1}{k} \binom{N-j}{d-k-1} 
\end{align}

We change the summation index by one, $i=j-1$, and then extend the sum up to $N-1$,
\begin{align}
\Sigma_1 
&= N \sum_{i=0}^{N-2} {\binom{i}{k} \binom{N-i-1}{d-k-1}} \nonumber \\
&= N \left[ \sum_{i=0}^{N-1} {\binom{i}{k} \binom{N-i-1}{d-k-1} - \binom{N-1}{k}\binom{0}{d-k-1}}\right].
\end{align}
The last term is zero as long as $d-k-1>0$, i.e.\ $k<d-1$. We can now apply a variant of the Vandermonde's convolution, $\sum_{i=0}^{l}\binom{l-i}{m}\binom{q+i}{n}=\binom{l+q+1}{m+n+1}$ \cite{Graham94} , on the first term and obtain for $k<d-1$ the result
$
\Sigma_1  
=N \binom{N}{d}.$
For the special case of $k=d-1$, we start from Eq. \eqref{eq27}, 
\begin{align}
\Sigma_1= N \sum_{j=1}^{N-1} \binom{j-1}{d-1} \binom{N-j}{0}
&= N \sum_{j=1}^{N-1} \binom{j-1}{d-1}.
\end{align}
Using the identity
$\sum_{j=1}^{N-1} \binom{j-1}{d-1} = \binom{N-1}{d}$,
we obtain 
$\Sigma_1 = N \binom{N-1}{d} = (N-d) \binom{N}{d}$. 
To summarize, we have for $\Sigma_1$
\begin{align}
\label{sum1}
\Sigma_1= \left\{ \begin{array}{ll}
N \binom{N}{d}  &\mbox{ for $0 \leq k < d-1$}\\
N \binom{N-1}{d} =  (N-d) \binom{N}{d} &\mbox{ for $k=d-1$}
       \end{array}. \right.
\end{align}
The second sum in Eq.\ \eqref{eq16} involving the additional factor $j$ can be rewritten as
\begin{align}
\label{eq19}
\Sigma_2 = \sum_{j=1}^{N-1} j \binom{j-1}{k} \binom{N-j}{d-k-1} = (k+1) \sum_{j=1}^{N-1} \binom{j}{k+1} \binom{N-j}{d-k-1},
\end{align}
where we have used $j \binom{j-1}{k}=(k+1)\binom{j-1}{k+1}$.
We again shift the summation index by one, $i=j-1$, and extend the sum up to $N-1$
\begin{align}
\Sigma_2 &= (k+1) \sum_{i=0}^{N-2}\left[\binom{i+1}{k+1} \binom{N-i-1}{d-k-1} \right] \nonumber \\
&=(k+1)\sum_{i=0}^{N-1}\left[\binom{i+1}{k+1} \binom{N-i-1}{d-k-1} \right] - (k+1)\left[\binom{N}{k+1}\binom{0}{d-k-1}\right].
\end{align}
The last term is zero for $k<d-1$. For the first term, we can apply the same variant of Vandermonde's convolution as above, $\sum_{i=0}^{l}\binom{l-i}{m}\binom{q+i}{n}=\binom{l+q+1}{m+n+1}$, 
and obtain 
\begin{align}
\label{eq:reskgeneral}
\Sigma_2 &= (k+1) \binom{N+1}{d+1}.
\end{align}
For $k=d-1$, we again start from Eq.\ \eqref{eq19}, which yields
\begin{align}
\Sigma_2 &= 
d \sum_{j=1}^{N-1} \binom{j}{d}\binom{N-j}{0}
= 
d \sum_{j=1}^{N-1} \binom{j}{d}
= d \binom{N}{d+1}.
\end{align}
We slightly rearrange these two results to a common binomial,
\begin{align}
\Sigma_2= \left\{ \begin{array}{ll}
(k+1)  \frac{N+1}{d+1} \binom{N}{d} &\mbox{ for $0 \leq k < d-1$} \\ 
 \frac{d}{d+1} (N-d) \binom{N}{d} &\mbox{ for $k=d-1$} 
       \end{array}. \right.
\end{align}
Combining these results with Eq.\ \eqref{sum1}, we obtain 
\begin{align}
\Sigma_1-\Sigma_2 =
 \binom{N}{d}  \frac{1}{d+1}
 \times
 \left\{ \begin{array}{ll}
N(d-k)-k-1 &\mbox{ for $0 \leq k < d-1$} \\ 
N-d &\mbox{ for $k=d-1$} 
       \end{array}. \right.
\end{align}
Note that these two expressions have the same form, such that we obtain a single expression
for $\Sigma_1-\Sigma_2$ or, equivalently, for Eq.\ \eqref{eq16},  
\begin{align}
\sum_{m=1}^{N-1}\sum_{j=1}^{m}   \binom{j-1}{k} \binom{N-j}{d-k-1}  = \Sigma_1-\Sigma_2 =
 \binom{N}{d}  \frac{N(d-k)-k-1}{d+1}.
\end{align}
Together with the common factor $\binom{N-1}{d-1}^{-1}$, we obtain
\begin{align}
\sum_{m=1}^{N-1}\sum_{j=1}^{m}\frac{\binom{j-1}{k} \binom{N-j}{d-k-1}}{\binom{N-1}{d-1}} 
=      \frac{N^2(d-k)-N(k+1)}{d(d+1)},
\end{align}
which is Eq.\ \eqref{eq:toprove2}.

In a similar way, the sums over $\pi_B$ can be solved. In that case, the special case is $k=0$ rather than
$k=d-1$, which also indicates the symmetry of the result.
For the sums over $\pi_B$, we obtain
\begin{align}
\sum_{m=1}^{N-1}\sum_{j=1}^{m}\frac{\binom{j}{k} \binom{N-j-1}{d-k-1}}{\binom{N-1}{d-1}} =
 \left\{ \begin{array}{ll}
\frac{N (N-d)}{d+1} &\mbox{ for $k = 0$} \\ \\
\frac{N (N+1) (d-k)}{d (d+1)} &\mbox{ for $1 \leq k \leq d-1$} 
       \end{array}. \right.
\end{align}

\section{Condition for the comparison of two strategies}
\label{appcond2}
The statement to prove is,
\begin{align}
 \sum_{j=1}^{N-1}(\pi_A-\pi_B) &=\frac{N}{d} \sum_{k=0}^{d-1}(a_k-b_k) +b_0-a_{d-1}.
\end{align}
As the $a_k$'s are contributed only by $\pi_A$ and the $b_k$'s only by $\pi_B$, we first need to show that,
\begin{align}
 \sum_{j=1}^{N-1}\pi_A &=\frac{N}{d}\sum_{k=0}^{d-1}a_k - a_{d-1},
\end{align}
with the payoffs from Eq.\ \eqref{payofffin}.
This holds for any choice of $a_k$'s.
Thus we only have to show that,
\begin{align}
\label{eq:toprove}
\frac{1}{\binom{N-1}{d-1}} \sum_{j=1}^{N-1} \binom{j-1}{k} \binom{N-j}{d-k-1} = \left\{ \begin{array}{ll}
\frac{N}{d}  &\mbox{ for $0 \leq k < d-1$}\\
 \frac{N}{d}-1 &\mbox{ for $k=d-1$} 
       \end{array}. \right.
\end{align}
The sum has been solved above, cf.\ Eq. \eqref{eq27}, where we have
shown that
$ \sum_{j=1}^{N-1} \binom{j-1}{k} \binom{N-j}{d-k-1} = \binom{N}{d}$ for  $0 \leq k < d-1$
and 
$ \sum_{j=1}^{N-1} \binom{j-1}{k} \binom{N-j}{d-k-1} = \frac{N-d}{N}\binom{N}{d}$ for  $k = d-1$.
Using the identity $\binom{N}{d} = \frac{N}{d} \binom{N-1}{d-1}$,
we directly obtain Eq.\ \eqref{eq:toprove}.

The equivalent condition for $\pi_B$ can be derived based on a similar argument.  As above, we have $k=0$ as the special case instead of $k=d-1$ in the equivalent of Eq.\ \eqref{eq:toprove},
\begin{align}
\frac{1}{\binom{N-1}{d-1}} \sum_{j=1}^{N-1} \binom{j}{k} \binom{N-j-1}{d-k-1} = \left\{ \begin{array}{ll}
\frac{N}{d}-1  &\mbox{ for $k=0$}\\
 \frac{N}{d} &\mbox{ for $0<k \leq d-1$} 
       \end{array}. \right.
\end{align}
